\documentclass[cameraready]{Interspeech}

\title{Assessing True Generalisability of Audio-Visual Speech Recognisers}

\author[affiliation={1,2}, orcid=0009-0009-3552-9944]{Zhaofeng}{Lin}
\author[affiliation={2,3}]{Stavros}{Petridis}
\author[affiliation={2,3}]{Maja}{Pantic}
\author[affiliation={1}]{Naomi}{Harte}

\address{
    $^1$ Trinity College Dublin, Ireland \\
    $^2$ Imperial College London, UK 
    $^3$ NatWest AI Research, UK
}

\email{linzh@tcd.ie}


\usepackage{comment}
\usepackage{booktabs}
\usepackage{multirow}
\usepackage{float}

\begin{document}

\maketitle

\begin{abstract}
Current Audio-Visual Speech Recognition (AVSR) models achieve near-perfect performance on the standard LRS3 benchmark, raising concerns of adaptive overfitting. To systematically assess true generalisability, we construct a highly controlled, unseen evaluation set subsampled from the massive MultiVSR dataset. Unlike standard out-of-distribution benchmarks, our subset strictly matches the acoustic, visual, and demographic distributions of the LRS3 test set. Evaluating five state-of-the-art architectures reveals a universal performance collapse, proving that current systems fail to generalise even under strictly aligned conditions. Through a fine-grained attribute analysis across seven factors, we isolate the specific drivers of this degradation. Furthermore, we uncover a profound lexical bias, expose distinct error patterns, and surprisingly reveal that audio-visual performance even lags behind audio-only settings. We release our matched test set for future benchmarking.
\end{abstract}

\section{Introduction}
For decades, Audio-Visual Speech Recognition (AVSR) research has sought to enhance conventional audio-only speech recognition by exploiting visual cues from the speaker's lip movements \cite{1230212, afouras2018deep}. In recent years, the AVSR field has experienced a rapid architectural evolution, advancing from supervised end-to-end networks \cite{sterpu2018attention, petridis2018end, ma2023auto} and self-supervised learning frameworks \cite{shi2022learning, haliassos2024unified} to the latest integrations with speech foundation models \cite{rouditchenko24_interspeech} and Large Language Models \cite{10889251}.

However, the evaluation of these recent, highly advanced architectures now relies almost entirely on a single benchmark: LRS3 \cite{afouras2018lrs3}. 
State-of-the-art (SoTA) methods are aggressively optimised for this specific dataset, and hence we are now witnessing severe performance saturation, with models achieving less than a 1\% Word Error Rate (WER) on clean speech and less than 5\% under 0 dB noise. Crucially, the LRS3 test set is exceptionally small, comprising only 0.9 hours of video against a 433-hour training corpus. This imbalance raises a critical question: does this near-perfect performance indicate a genuine ability to recognise audio-visual speech, or does it merely reflect strong overfitting to the LRS3 distribution?

A robust AVSR system should generalise well to unseen data. 
Researchers in the broader computer vision field have assessed the generalisability of image classifiers by constructing new test sets that strictly replicate the data curation procedures of the original ImageNet dataset \cite{recht2019imagenet}. 
Their methodology isolates whether performance degradation stems from a true distribution gap, or extensive adaptation to the original dataset's specific characteristics.
Similar efforts have recently emerged in Visual Speech Recognition (VSR), challenging the LRS3 benchmark in that domain \cite{djilali2024vsr}. However, while the resulting WildVSR dataset is valuable for the visual-only community, it completely omits audio tracks. As a result, it cannot be used to evaluate AVSR systems, leaving a critical gap in the field.

While replicating the data curation process has proven to be effective in assessing generalisability, constructing a new audio-visual dataset of sufficient scale presents extreme challenges. Therefore, instead of following the LRS3 creation process like WildVSR, we present an alternative, rigorous method in this paper.
We subsample a highly controlled, unseen LRS3-like evaluation set from the massive MultiVSR dataset \cite{10890395}, ensuring this subset strictly matches the LRS3 test distribution.
We then systematically investigate the generalisability of state-of-the-art AVSR systems. 
Specifically, we evaluate the capacity of these models to maintain their performance on data that shares the same distribution of acoustic and visual factors as LRS3. 
Furthermore, we isolate and analyse the specific factors that cause the most severe degradation when performance inevitably drops.
Our key findings and contributions are summarised as follows:
\begin{itemize}
    \item We demonstrate a severe performance drop across all evaluated SoTA models when tested on the unseen LRS3-like set.
    \item We conduct detailed attribute analysis across 7 factors (duration, age, gender, skin tone, head pose, SNR, speech rate) to identify the most influential causes of performance drop.
    \item We uncover a profound lexical bias within current models. By isolating vocabulary effects, we reveal a performance gap between words shared with the original LRS3 test set and those outside its narrow lexical distribution. 
    \item We compare the audio-only, video-only and audio-visual settings, and surprisingly find audio-visual performance is worse than audio-only for most models.
    \item We find models have very distinct patterns in substitution, deletion, and insertion errors.  
    \item We release our test set MultiVSR2LRS3 (MV2LRS3) and all extracted metadata.\footnote{\url{https://github.com/chaufanglin/mv2lrs3}}
\end{itemize}

\section{Constructing a Matched Test Set}
Prior efforts in the broader computer vision field \cite{recht2019imagenet} and VSR \cite{djilali2024vsr} provide valuable inspiration for assessing the true generalisation of machine learning models.
However, these foundational studies mainly analyse distribution shifts and the resulting performance degradation from a broad, statistical machine learning perspective.
In the highly complex domain of audio-visual speech, a purely statistical approach is insufficient.
To understand true generalisability of AVSR, we must ground our distribution matching within a speech context, specifically focusing on factors that will impact speech recognition performance.

In this paper, we construct a new evaluation set strictly matching the demographic, acoustic and visual distributions of the LRS3 test set. 
We derive this subset from a recently created, massive lip-reading speech dataset, MultiVSR \cite{10890395}. Specifically, we establish the LRS3 Test set as our reference distribution and utilise MultiVSR as our candidate pool. 
Our primary objective is to subsample this candidate pool to extract a distributionally aligned evaluation set, which we term the \textbf{MultiVSR2LRS3 (MV2LRS3)}. We briefly outline the contents of the two datasets in the next section, and then present our rationale and approach in distribution matching.

\subsection{Datasets and Preprocessing}
\noindent\textbf{LRS3:} 
The LRS3-TED dataset \cite{afouras2018lrs3}, known as LRS3, comprises over 400 hours of video sourced from 5594 TED and TEDx talks in English, downloaded from YouTube. LRS3 contains $\sim$150K utterances (439 hours). The breakdown of this dataset includes 408 hours in the pretraining set, 30 hours in the train-val set, and 0.9 hours (1321 utterances) in the test set. 

\noindent\textbf{MultiVSR:} 
MultiVSR \cite{10890395} is a large-scale multilingual dataset, comprising nearly 12,000 hours of video data. It was sourced from $\sim$200K publicly available YouTube videos, and the video IDs are exactly the same as the AVSpeech dataset \cite{ephrat2018looking}. While AVSpeech only uses a few seconds from the videos, MultiVSR uses as much as it can from the source videos. 
It has 13 languages, and in our experiment, we only used the English part.
However, the authors only designate a train and validation set for English, so we use the validation set for all our experiments. 

\noindent\textbf{Pre-processing:}
While MultiVSR provides pre-cropped face tracking videos and corresponding transcriptions, it lacks the audio tracks as it was originally targeted at VSR. To construct a complete evaluation set, we utilised the provided YouTube IDs to download the source media and extract the audio tracks.
We then followed the Auto-AVSR pre-processing pipeline to detect and crop mouth regions of interest (ROIs) from the videos. 
Following the LRS3 curation process, we divided the videos into individual sentence utterances based on transcript punctuation marks (full stops, commas, question marks, and exclamation marks). 
Finally, since the MultiVSR transcriptions were machine-generated using the Whisper Large-v3 model via WhisperX \cite{bain23_interspeech}, we conducted a manual verification on a one-hour subsample, confirming a low WER of only 2.3\%.

\subsection{Choice of Potential Factors}

Established research in audio-only speech recognition demonstrates a clear performance gap linked to demographic and acoustic factors. Challenges include age, gender, native accents and dialect \cite{feng2024towards}, influence of L1 for non-native speakers \cite{liu2022towards}, acoustic SNR \cite{6732927}, and speech rate \cite{kumar2025performance}.
Meanwhile, VSR literature establishes that visual factors like head pose \cite{djilali2024vsr}  and lexical factors like vocabulary \cite{gimeno2025contributions} also strongly influence performance. Studies in the facial‑recognition domain indicate that image‑based encoders exhibit performance biases linked to skin tone \cite{cavazos2020accuracy}. It is plausible that such biases may transfer into AVSR systems, as they rely on similar visual representations. 

To isolate relevant factors, and construct our controlled MV2LRS3 set, we identified seven key metadata attributes that dictate the distribution of the LRS3 Test set. We extract values for these attributes from all samples in both the baseline LRS3 Test set and the MultiVSR English Validation set, enabling precise distribution matching.

\textbf{Age and Gender}: We use the open-source Uniface toolkit\footnote{\url{https://github.com/yakhyo/uniface}} to estimate the demographic factors of age and gender directly from the video frames.

\textbf{Apparent Skin Tone}: We employ the open-source Stone skin tone classifier \cite{rejon2023classification} to detect skin tone on the Monk Skin Tone (MST) scale \cite{Monk_2019}. We specifically note that this metric captures the apparent surface colour and, whilst closely linked to ethnicity, is also influenced by lighting conditions. This makes it a highly relevant factor for evaluating robustness in varied, real-world environments.

\textbf{Head Pose}: We use 6DRepNet \cite{9897219} to estimate the pitch, yaw, and roll of the speaker's head per video frame. Among these, we choose yaw (left-right rotation) as our primary spatial metric, as it impacts AVSR performance most \cite{8756582}.

\textbf{Signal-to-Noise Ratio (SNR)}: To measure acoustic degradation and environmental interference, we compute the SNR of the audio streams using the WADA-SNR algorithm \cite{kim08e_interspeech}.

\textbf{Duration and Speech Rate}: Finally, we calculate the utterance duration and the speech rate (measured in number of words per second) in each sample.

\subsection{Distribution Matching}

We construct a controlled subset of the MultiVSR English validation set that strictly mirrors the LRS3 reference distribution.
We achieve this through a multidimensional k-Nearest Neighbour (kNN) \cite{cunningham2021k} matching strategy.

First, we represent every video utterance in both corpora as an eight-dimensional feature vector. 
This comprises age, gender, skin tone, SNR, video duration, speech rate, and head pose. As yaw is estimated per frame, we represent it at the video level using both its mean and standard deviation, accounting for the eighth dimension.
These attributes exist on fundamentally different scales (some continuous and others categorical) and hold varying degrees of domain significance. Thus, an unweighted distance metric would allow certain features to dominate the selection process disproportionately. 
To avoid this, we implement an empirical feature weighting strategy. 
We iteratively apply specific scalar weights to each feature to optimise the covariate balance: Duration (100), Age (100), Gender (50), Skin Tone (40), Yaw Mean (100), Yaw Std (50), SNR (70), and Speech Rate (40). We validate the suitability of these heuristic weights by visually inspecting the plotted distribution of each factor to confirm a close fit between the matched set and LRS3 Test set.

\begin{figure}[htb]
  \centering
  \includegraphics[width=1\linewidth]{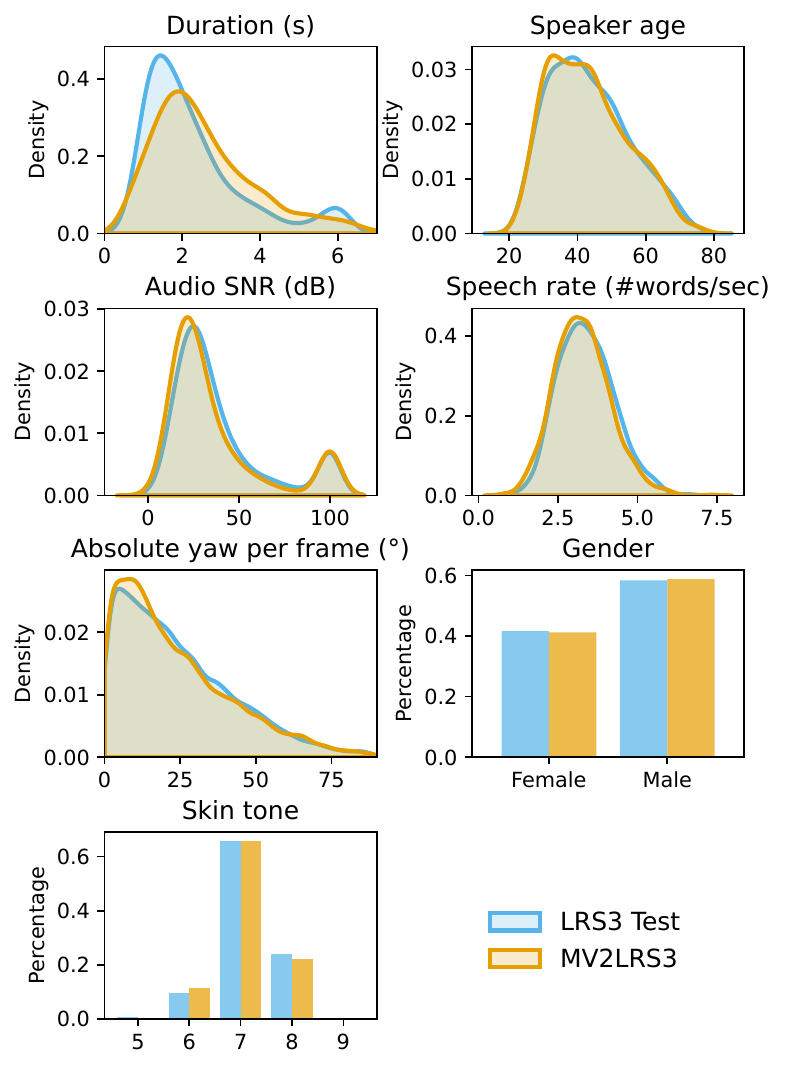}
  \caption{Distribution of all 7 factors on the LRS3 Test set and MV2LRS3 set.}
  \label{fig:base_dist}
\end{figure}

For every weighted utterance vector in the LRS3 test set, we calculate the Euclidean distance to all vectors in the MultiVSR candidate pool and identify the five closest matching samples.
To introduce controlled statistical variance while maintaining the strict reference distribution, we uniformly sample one utterance at random from these five candidates.
We repeat this stochastic selection process five times to generate five variations of the controlled subset, collectively termed the \textbf{MV2LRS3}.
Consequently, the obtained MV2LRS3 closely aligns with the distribution of the LRS3 Test set across all attributes.
Figure \ref{fig:base_dist} shows the distributions of all attributes for both the LRS3 Test set and the MV2LRS3.

\section{Evaluated Models}

We select five state-of-the-art models representing the rapid architectural evolution of AVSR over recent years. These encompass fully-supervised end-to-end models, self-supervised models fine-tuned on LRS3, and the latest architectures integrating speech foundation models and Large Language Models:
\begin{itemize}
    \item \textbf{AV-HuBERT} \cite{shi2022learning}: A self-supervised learning framework that pre-trains on unlabelled audio-visual data to learn robust representations and then fine-tunes on the LRS3 benchmark. 

    \item \textbf{Auto-AVSR} \cite{ma2023auto}: A fully-supervised, end-to-end AVSR architecture trained on massive labelled datasets, leveraging Whisper to generate pseudo-labels for unlabelled corpora such as AVSpeech \cite{ephrat2018looking} and VoxCeleb2 \cite{chung18b_interspeech}.
    
    \item \textbf{USR} \cite{haliassos2024unified}: A unified teacher-student framework capable of ASR, VSR, and AVSR within a single architecture, which pre-trains on unlabelled audio-visual data before applying semi-supervised fine-tuning via pseudo-labels.

    \item \textbf{Whisper-Flamingo} \cite{rouditchenko24_interspeech}: An audio-visual model integrating AV-HuBERT visual features into Whisper \cite{radford2023robust}, an audio-only foundation model trained on 680k hours of speech. 
    
    \item \textbf{Llama-AVSR} \cite{10889251}: A multimodal framework that integrates pre-trained encoders (Whisper for audio, AV-HuBERT for visual) into a Large Language Model to jointly perform ASR, VSR, and AVSR. 
\end{itemize}

\begin{table*}[ht]
\caption{Summary of model training data alongside AVSR performance (WER \% and ranking) on the LRS3 Test and MV2LRS3 sets. MV2LRS3 results represent the mean and standard deviation over 5 runs.}
\label{tab:master_results}
\centering
\begin{tabular}{l l c c cccc} 
\toprule
\multirow{2}{*}{\textbf{Model}} & \multirow{2}{*}{\textbf{Training Data}} & \multirow{2}{*}{\begin{tabular}[c]{@{}c@{}}\textbf{Unlabelled}\\\textbf{hours}\end{tabular}} & \multirow{2}{*}{\begin{tabular}[c]{@{}c@{}}\textbf{Labelled}\\\textbf{hours}\end{tabular}} & \multicolumn{2}{c}{\textbf{LRS3}} & \multicolumn{2}{c}{\textbf{MV2LRS3}} \\ 
\cmidrule(l){5-6} \cmidrule(l){7-8}
 &  &  &  & \textbf{WER} & \textbf{Rank} & \textbf{WER} & \textbf{Rank} \\ 
\midrule
AV-HuBERT & LRS3, VoxCeleb2 & 1,326 & 433 & 1.50 & 5 & 23.5 $\pm$ 0.6 & 5 \\
Auto-AVSR & LRW, LRS3, VoxCeleb2, AVSpeech & - & 3,448 & 0.95 & 3 & 14.0 $\pm$ 0.3 & 1 \\
USR & LRS3, VoxCeleb2 & 1,326 & 433 & 1.10 & 4 & 21.5 $\pm$ 0.3 & 4 \\
Whisper-F. & LRS3, VoxCeleb2 & - & 1,759 & 0.86 & 2 & 18.6 $\pm$ 1.0 & 3 \\
Llama-AVSR & LRS3, VoxCeleb2 & - & 1,759 & 0.77 & 1 & 16.5 $\pm$ 0.4 & 2 \\
\bottomrule
\end{tabular}
\end{table*}

\section{MV2LRS3 Set Evaluation}
Having constructed a subset strictly matched to the LRS3 distribution across all seven factors, we evaluate the WER of the five chosen models. Table \ref{tab:master_results} details their performance and relative rankings on both LRS3 benchmark and our MV2LRS3 set.

\subsection{Overall Performance}
Our evaluation reveals a severe degradation in WER across all SoTA systems. On the LRS3 Test set, all models demonstrate near-perfect performance, achieving a WER below 1.5\%, with top-performing systems dropping below 1\%.
However, despite our MV2LRS3 set sharing a closely-aligned statistical distribution across all seven metadata factors, no model sustains this accuracy. Performance universally collapses, with the lowest error rate surging to 14.0\%, and the highest reaching 23.5\%.

Beyond the absolute performance drop, the relative ranking of the models shifts in this new environment. Llama-AVSR, the best-performing model on LRS3 (0.77\%), falls to second place, while Auto-AVSR climbs from third place to take the top rank (14.0\%). 
However, it is important to remember that Auto-AVSR was trained on AVSpeech. MultiVSR shares this exact video source, thus the model's higher ranking might benefit from domain familiarity rather than true generalisability.

\subsection{Linear Fit Analysis}
\begin{figure}[ht]
  \centering
  \includegraphics[width=0.8\linewidth]{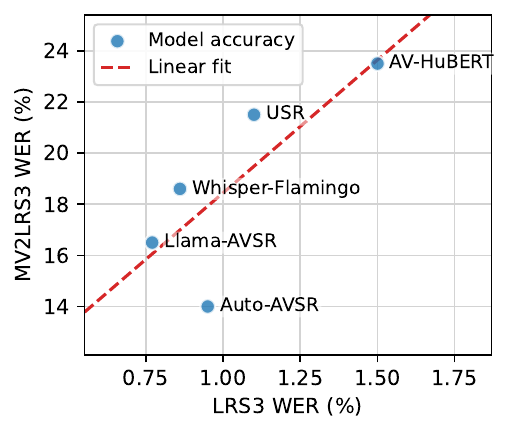}
  \caption{Model performance on LRS3 Test v.s. MV2LRS3 set.}
  \label{fig:linear_fit}
\end{figure}

Figure \ref{fig:linear_fit} shows the relationship between WER on the MV2LRS3 and the LRS3 Test set.
We observe that the performance degradation can be approximated by a linear function:
\begin{equation}
    WER_{MV2LRS3} = 10.4 \times WER_{LRS3} + 8.1 
\label{eq:linear}
\end{equation}

The steep slope of 10.4 demonstrates a high degree of sensitivity; small performance variations on the LRS3 benchmark are amplified around tenfold on the MV2LRS3 set. We can compare this to established generalisability literature for image classification and VSR \cite{recht2019imagenet, djilali2024vsr}, where linear fit slopes typically range between 1.0 and 2.0. 
In contrast, our observed slope of 10.4 is substantially steeper than theirs. 

Within this trend, Auto-AVSR is the only model sat below the linear fit. This placement indicates that it maintains better robustness than the expected average, allowing it to overtake Llama-AVSR and Whisper-Flamingo in our MV2LRS3 set rankings. 

\subsection{Validating with a Scaled 10x Subset}

To validate the robustness of our findings, we scale our evaluation by constructing a matched subset 10 times larger than the original MV2LRS3 set, increasing the total test duration from 0.9 hours to approximately 10 hours, termed the \textbf{10x set}. 
Scaling the selection pool forces the algorithm to accept candidates with greater distances, and achieving an exact distributional fit becomes more difficult. However, while the resulting distribution slightly relaxes, it remains closely aligned with the LRS3 Test set. 
The resulting distribution of the 10x set is shown in Figure \ref{fig:difficulties}.
This expansion allows us to observe whether the performance degradation and the changes in model rankings persist when evaluated on a substantially larger test set.

The results for this expanded evaluation are shown in Table \ref{tab:10x}. We observe that the fundamental degradation in performance persists across all models. 
While the absolute magnitude of this drop is slightly smaller on the 10x set, the performance ranking of the models on the 10x set is still identical to the ranking observed on the MV2LRS3 set. 
This consistent ranking across a substantially larger test set confirms that our findings reflect the true generalisability of the SoTA models.
To avoid confusion, in the following experiments, we mainly use the MV2LRS3 set and will explicitly mention if using the 10x set.

\begin{table}[htb]
\centering
\caption{WER (\%) of SoTA models on MV2LRS3 (mean over 5 runs) and 10x set (single run).}
\label{tab:10x}
\begin{tabular}{l|cc|cc} 
\toprule
\textbf{Model}   & \textbf{MV2LRS3} & \textbf{Rank} & \textbf{10x} & \textbf{Rank}  \\ 
\midrule
AV-HuBERT  & 23.5  & 5 & 21.2   & 5  \\
Auto-AVSR  & 14.0  & 1  & 12.9   & 1  \\
USR  & 21.5  & 4  & 19.4   & 4  \\
Whisper-F. & 18.6  & 3  & 16.0   & 3  \\
Llama-AVSR & 16.5  & 2  & 13.5   & 2  \\
\bottomrule
\end{tabular}
\end{table}

\section{Isolating the Impact of Each Factor}

\subsection{Leave-one-out Attribute Analysis}
\label{sec:ablation}
\begin{table*}[ht]
\centering
\caption{Leave-one-out attribute analysis evaluating the isolated impact of individual factors on model performance. 
Results denote WER and the relative percentage change (\% Rel. Change) compared to the MV2LRS3 set. 
Each column represents a subset where the indicated factor was excluded from the matching constraints, allowing it to vary naturally while the remaining six factors were strictly matched to the LRS3 distribution. MV2LRS3 and leave-one-out results represent the mean over 5 runs.}
\label{tab:ablation}
\setlength{\tabcolsep}{3pt}
\begin{tabular}{l c c ccccccc} 
\toprule
\multirow{2}{*}{\textbf{Model}} & \multirow{2}{*}{\textbf{LRS3}} & \multirow{2}{*}{\textbf{MV2LRS3}} & \multicolumn{7}{c}{\textbf{Subsets excluding one factor [WER (\% Rel. Change)]}} \\ 
\cmidrule(lr){4-10}
 & & & \textbf{Duration} & \textbf{Age} & \textbf{Gender} & \textbf{Skin tone} & \textbf{Yaw} & \textbf{SNR} & \textbf{Speech rate} \\ 
\midrule
AV-HuBERT & 1.50 & 23.5 & 15.8 (-33\%) & 22.8 (-3\%) & 22.3 (-5\%) & 21.9 (-7\%) & 22.3 (-5\%) & 23.1 (-2\%) & 22.5 (-5\%) \\
Auto-AVSR & 0.95 & 14.0 & 10.6 (-24\%) & 14.2 (+2\%) & 13.8 (-1\%) & 13.5 (-4\%) & 13.4 (-4\%) & 13.5 (-4\%) & 14.0 (-) \\
USR & 1.10 & 21.5 & 13.7 (-36\%) & 21.1 (-2\%) & 20.5 (-5\%) & 20.4 (-5\%) & 20.1 (-6\%) & 21.1 (-2\%) & 20.7 (-4\%) \\
Whisper-F. & 0.86 & 18.6 & 9.9 (-47\%)  & 18.6 (-)    & 18.2 (-2\%) & 16.9 (-9\%) & 17.1 (-8\%) & 18.1 (-3\%) & 19.4 (-5\%) \\
Llama-AVSR& 0.77 & 16.5 & 16.8 (+2\%)  & 14.1 (-14\%)& 14.3 (-13\%)& 15.0 (-10\%)& 14.2 (-14\%)& 14.3 (-13\%)& 17.5 (-6\%) \\
\bottomrule
\end{tabular}
\end{table*}

During the construction of our controlled MV2LRS3 set, we identified 7 distinct attributes with the potential to influence AVSR performance. Guided by established research concerning demographic and acoustic bias in the ASR field \cite{liu2022towards, feng2024towards}, we hypothesise that specific factors may have a greater impact on the overall performance degradation.

To find the cause of the performance drop on the MV2LRS3 set, we conduct a leave-one-out attribute analysis on the MV2LRS3 set. We repeat our k-Nearest Neighbour matching to build new subsets, but we exclude exactly one factor each time. In every new set, six factors remain strictly matched to the LRS3 distribution, while the seventh factor is allowed to vary in an unconstrained manner. By testing the models on these subsets and comparing the WER to our original MV2LRS3 set, we can more clearly isolate the impact of each individual factor.
Table \ref{tab:ablation} shows the full results for this experiment. 

\noindent\textbf{Duration.}
When we remove the duration constraint, the subset naturally includes much longer samples, reaching up to 20 seconds compared to the strict seven-second limit of LRS3. This extended duration dramatically improves performance for most models. For Whisper-Flamingo, the WER drops to 9.9\%, representing a 47\% relative improvement compared to the MV2LRS3 set. 
AV-HuBERT, Auto-AVSR, and USR are also substantially affected, and the WER relative improvement varies from 24\% to 36\%.
This suggests that these specific architectures benefit heavily from the expanded linguistic context available in longer video sequences.

However, Llama-AVSR is a distinct outlier in this scenario, performing slightly worse than on the MV2LRS3 set.
Our investigation reveals this is caused by a hard architectural bottleneck rather than the data itself. Since the LRS3 Test set features exclusively short utterances, Llama-AVSR was highly optimised for this specific environment with a strict maximum output limit of 32 tokens. When forced to process longer utterances, this token limit artificially truncates the output. Furthermore, our attempts to increase this token allowance resulted in the model hallucinating, exposing a severe limitation in its underlying LLM integration.

\noindent\textbf{Other factors.}
The impacts of the remaining factors are less dramatic but still notable. When skin tone or yaw are left unmatched, Llama-AVSR and Whisper-Flamingo show WER improvements between 8\% and 14\%, while the other models show a smaller relative improvement. Conversely, when age, gender, SNR, or speech rate are unmatched, most models show less than a 5\% relative difference from the MV2LRS3 set. Llama-AVSR remains the notable exception; it exhibits a 6\% relative difference for speech rate, while its variation climbs to over 10\% for age, gender, and SNR.

Ultimately, this attribute analysis highlights clear differences in model robustness. Auto-AVSR proves to be the most stable and reliable model across all varying factors. In contrast, Llama-AVSR is highly sensitive, behaving unpredictably when forced outside the specific format of its LRS3 training data.

\subsection{Dive into the Influential Factors}

\subsubsection{Duration}
In Section \ref{sec:ablation}, we found that the duration heavily impacts the performance, and the pattern of the models varies. 
We separate the data into two discrete bins: 0-3 seconds and 3-7 seconds, and we match all other statistical factors with the LRS3 Test set to guarantee that any performance differences are strictly driven by duration. 
Table \ref{tab:bin} shows the performance of all AVSR models on the two duration bins.

Short utterances (0-3 seconds) are considerably harder for all models to recognise. The error rates drop dramatically when the models are given longer utterances. 
For instance, WER of Whisper-Flamingo falls from 22.4\% on short clips to just 8.3\% on longer clips. This indicates that some architectures rely heavily on extended contextual information to accurately decode speech, struggling when forced to predict from short clips.

However, Auto-AVSR presents a strong exception. It achieves the lowest WER for short utterances (15.0\%), and the performance gap between the short and long bins is less than 5\%. This temporal stability suggests that Auto-AVSR does not depend as heavily on long-term linguistic context.

\subsubsection{Yaw}
Results in Section \ref{sec:ablation} demonstrate that head yaw is an influential factor, although not as dominant as duration. This aligns with recent findings in VSR \cite{djilali2024vsr}, which report that high variability in head pose severely degrades the ability to recognise speech.
To test visual robustness, we group the data into three angular bins: 0-30 degrees (mostly frontal), 30-60 degrees (partial profile), and 60-90 degrees (extreme profile). Table \ref{tab:bin} shows the performance of all AVSR models on the three yaw bins.

The result reveals a clear trend: most models experience a slight performance drop at extreme angles. 
Models like AV-HuBERT and Whisper-Flamingo perform worse in the 60-90 degree bin, likely because the extreme profile view physically hides lip movements from the visual encoder. 

However, Llama-AVSR is a notable exception. 
Its error rate remains highly stable across all three angles, shifting only from 14.2\% to 14.1\%. This stability is particularly unexpected because Llama-AVSR uses the exact same AV-HuBERT visual encoder as Whisper-Flamingo and the AV-HuBERT AVSR model. 
Since they share the same visual encoder but display completely different behaviour at extreme angles, this suggests that Llama-AVSR's backend language model is better at compensating for degraded visual inputs using strong linguistic context.

\begin{table}[ht]
\centering
\caption{WER (\%) on the binned subsets; Duration: 0-3 seconds and 3-7 seconds; Yaw: 0-30 degrees, 30-60 degrees, and 60-90 degrees.}
\label{tab:bin}
\begin{tabular}{lccccc} 
\toprule
\multirow{2}{*}{\textbf{Model}} & \multicolumn{2}{c}{\textbf{Duration}} & \multicolumn{3}{c}{\textbf{Yaw}}                 \\ 
\cmidrule(lr){2-3} \cmidrule(lr){4-6}
                                & \textbf{0-3} & \textbf{3-7}           & \textbf{0-30} & \textbf{30-60} & \textbf{60-90}  \\ 
\midrule
AV-HuBERT  & 26.2  & 16.0 & 22.4  & 20.8  & 23.6  \\
Auto-AVSR  & 15.0  & 10.4 & 13.5  & 13.0  & 15.1  \\
USR  & 24.0  & 14.6  & 20.5  & 19.0  & 21.4  \\
Whisper-F.  & 22.4  & 8.3  & 16.8  & 18.8  & 20.4 \\
Llama-AVSR  & 19.9  & 7.7  & 14.2  & 13.9  & 14.1  \\
\bottomrule
\end{tabular}
\end{table}

\subsection{Characterising Hard and Easy Samples for all AVSR}

\begin{figure}[htb]
  \centering
  \includegraphics[width=1\linewidth]{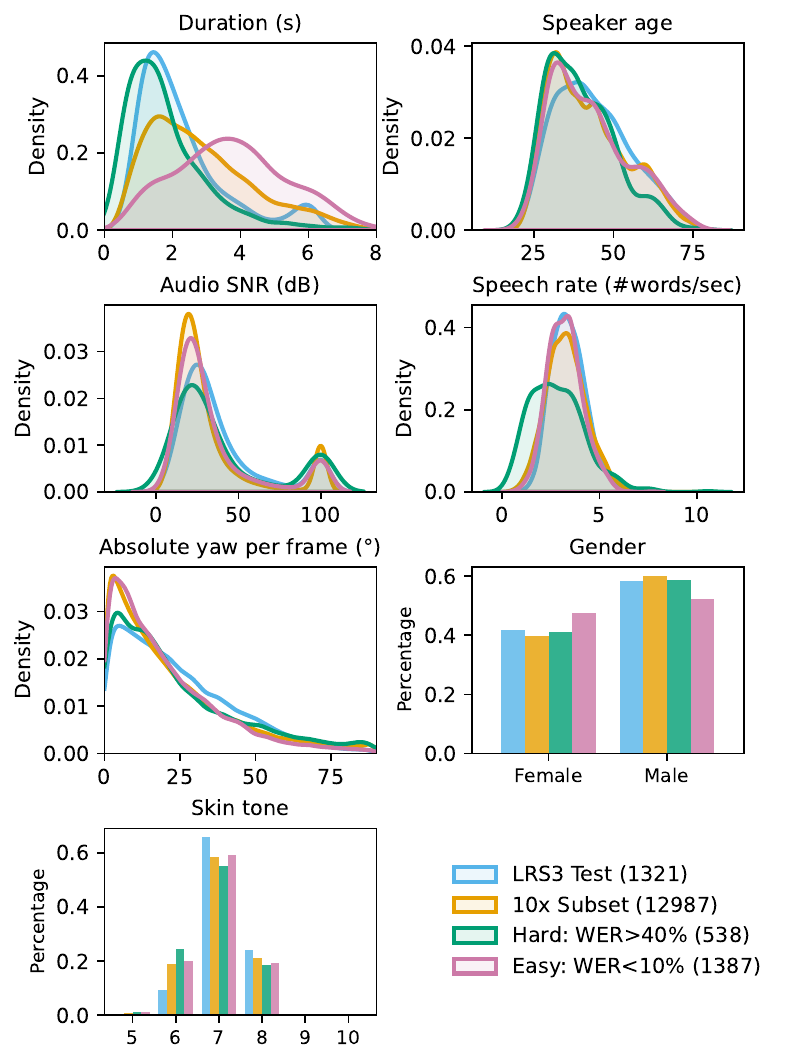}
  \caption{Distribution of all 7 factors on the LRS3 Test set, 10x set, Difficult set and Easy set. Numbers in the bracket denotes number of samples in the set.}
  \label{fig:difficulties}
\end{figure}

To identify the universal failure modes of AVSR systems, rather than model-specific bottlenecks, we conduct a distributional analysis of performance extremes. 
Profiling edge cases requires statistical significance, thus we isolate two distinct subsets from our expanded \textbf{10x set} rather than the original MV2LRS3 set. We designate an \textbf{Easy} subset (comprising samples where all evaluated models achieve a WER below 10\%) and a \textbf{Hard} subset (comprising samples where all models suffer a WER above 40\%). 
Figure \ref{fig:difficulties} presents the complete distributional comparison of all isolated attributes across the Easy and Hard subsets, alongside the LRS3 Test and 10x sets.

\noindent\textbf{Duration.}
Duration remains a critical factor in model performance.
The duration of the Easy subset exhibits a distinct, roughly normal distribution centred around 4 seconds. 
This suggests that current architectures are highly optimised for processing moderately sized context windows. 
In contrast, the Hard subset aligns more closely with the broader LRS3 distribution, or even shifts slightly towards shorter durations, confirming shorter utterances are generally more challenging for AVSR. 

\noindent\textbf{Speech rate.}
Counter-intuitively, our results reveal that samples within the Hard subset consistently demonstrate a lower overall speech rate compared to the Easy subset. 
While rapid speech is documented as a failure mode in some ASR research \cite{kumar2025performance}, this anomaly highlights a complex confounding variable in AVSR evaluation: it remains an open question whether excessively slow speech introduces inherent multimodal alignment difficulties, or if the models simply struggle because slow speech deviates so heavily from their primary training data, i.e. distributional familiarity.

\noindent\textbf{Yaw.}
The Hard subset contains less frontal yaw than the Easy subset. 
This aligns with the hypotheses and visual-only findings detailed in recent VSR literature \cite{djilali2024vsr}, that as the speaker's face turns away from the camera, the occlusion of lip movement causes systemic failures. 

\noindent\textbf{Others.}
Beyond the primary factors, the remaining profiles align with established demographic trends. For instance, the Easy subset contains a higher proportion of female speakers compared to the Hard set, which precisely aligns with common acknowledgements in ASR bias research \cite{liu2022towards, feng2024towards}.

\section{Impact of the Vocabulary}

Vocabulary divergence is a critical factor influencing AVSR performance, and it is important to isolate its specific impact. Evaluating this is complex because the SoTA models rely on vastly different training corpora. While some are fine-tuned exclusively on LRS3, others incorporate pseudo-labels from datasets like AVSpeech and VoxCeleb2. Furthermore, models leveraging Whisper as a foundational backbone have a non-transparent pre-training lexicon, making it impossible to determine their exact exposure to specific words.

To investigate the impact of the vocabulary, we categorise the MV2LRS3 set vocabulary based on its intersection with LRS3 Test set.
Let $V_\text{LRS3 Test}$ represent the complete vocabulary of the LRS3 Test set, and $V_\text{MV2LRS3}$ represent the vocabulary of our MV2LRS3 set.
The shared vocabulary present in both sets is defined as $V_\text{share} = V_\text{MV2LRS3} \cap V_\text{LRS3 Test}$.
The vocabulary present in the MV2LRS3 set but absent from the LRS3 Test set is defined as $V_\text{diff} =  V_\text{MV2LRS3} - V_\text{LRS3 Test}$.

\begin{figure}[tb]
  \centering
  \includegraphics[width=0.85\linewidth]{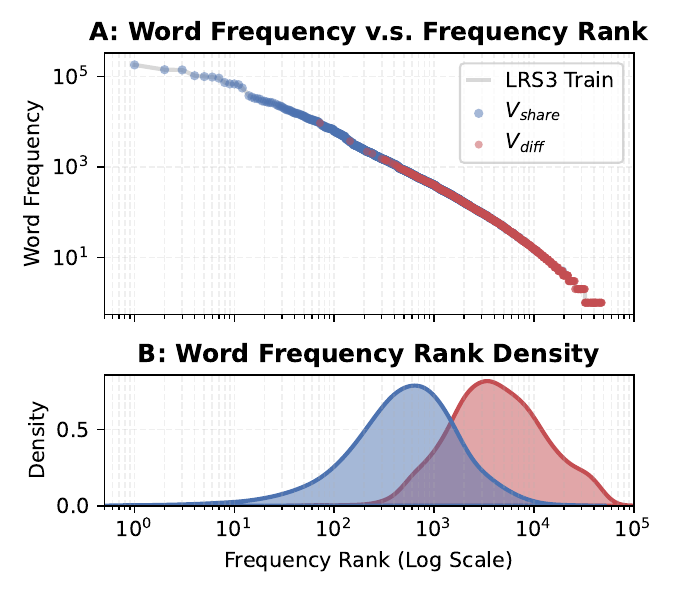}
  \caption{Distribution of the $V_\text{share}$ and $V_\text{diff}$ sets across the Zipfian frequency curve of the LRS3 training vocabulary.}
  \label{fig:zipf}
\end{figure}

Figure \ref{fig:zipf} visualises the distribution of the $V_\text{share}$ and $V_\text{diff}$ sets against the Zipfian frequency curve \cite{zipf2013psycho} of the LRS3 training vocabulary. 
Note that 883 out of 930 words in $V_\text{share}$ appear in LRS3 training vocabulary, and 1361 out of 1547 words in $V_\text{diff}$ appear in LRS3 training vocabulary.
As illustrated, $V_\text{share}$ is dominated by highly frequent words, whereas $V_\text{diff}$ is heavily skewed towards the long tail. This indicates a representation disparity: the models are exposed to the $V_\text{share}$ vocabulary potentially more often during training than the comparatively rare words in $V_\text{diff}$.

We calculate the Individual Word Error Rate (IWER) \cite{goldwater2010words} of the $V_\text{LRS3 Test}$ in LRS3 Test and $V_\text{share}$ and $V_\text{diff}$ sets in the MV2LRS3 set. 
The IWER of a set is calculated as the sum of substitution and deletion errors divided by the total occurrences of the set of words, defined inline as $IWER(V) = (S+D) / (H+S+D)$, where $S$ represents substitutions, $D$ represents deletions, and $H$ represents correct hits. Insertion errors are excluded here due to the difficulty in attributing them to specific words. 

\begin{table}[t]
\centering
\caption{IWER (\%) comparison between the LRS3 Test vocabulary ($V_{\text{LRS3 Test}}$) and the partitioned MV2LRS3 set vocabularies: $V_{\text{share}}$ and $V_{\text{diff}}$. $\Delta$ shows the IWER difference between $V_{\text{share}}$ and $V_{\text{diff}}$.}
\label{tab:vocab}
\begin{tabular}{lcccc} 
\toprule
\multirow{2}{*}{\textbf{Model}} & \multirow{2}{*}{$V_\text{LRS3 Test}$} & \multicolumn{3}{c}{\textbf{MV2LRS3}} \\ 
\cmidrule(l){3-5}
 &  & $V_\text{share}$ & $V_\text{diff}$ & $\Delta$ \\ 
\midrule
AV-HuBERT & 1.3 & 19.0 & 28.7 & 9.7 \\
Auto-AVSR & 0.8 & 8.5 & 19.9 & 11.4 \\
USR & 0.9 & 17.7 & 21.8 & 4.1 \\
Whisper-F. & 0.6 & 14.8 & 17.8 & 3.0 \\
Llama-AVSR & 0.6 & 9.1 & 15.2 & 6.1 \\
\bottomrule
\end{tabular}
\end{table}
\subsection{Results on the Shared Vocabulary}
Our results are shown in Table \ref{tab:vocab}. 
First, we observe a clear performance gap: the IWER on shared words ($V_{\text{share}}$) is substantially higher than on the $V_{\text{LRS3 Test}}$ set. 
Because our MV2LRS3 set strictly matched the seven primary acoustic and visual factors of the LRS3 Test set, this degradation indicates that there might be some other undiscovered factors affecting the performance.
For instance, Auto-AVSR and Llama-AVSR maintain a $V_{\text{share}}$ error rate below 10\%, whereas AV-HuBERT struggles severely, reaching 19\%.

\subsection{Insight from the Performance Gap}
Second, we compare the IWER across our two vocabulary sets, defining the performance gap between $V_\text{share}$ and $V_\text{diff}$ as $\Delta$. Before analysing the performance difference, we must explicitly define the boundaries of $V_{\text{diff}}$. 
Because this subset only excludes words found in the LRS3 Test set, it still contains vocabulary present in the broader LRS3 Training corpus (as shown in Fig. \ref{fig:zipf}) or massive external pre-training datasets. 
Therefore, we do not treat $V_{\text{diff}}$ as out-of-vocabulary. 
However, the words in $V_{\text{diff}}$ are rarer than words in $V_\text{share}$ within the LRS3 training vocabulary, potentially making them harder for the recognition.
Also, it serves as a critical measure of out-of-benchmark vocabulary: words that fall outside the highly expected linguistic distribution of the LRS3 evaluation process.

When examining the models, Whisper-Flamingo shows a minimal $\Delta$ between $V_{\text{share}}$ and $V_{\text{diff}}$.
Since Whisper-Flamingo uses Whisper as its backbone and Whisper is trained on 680k hours of data, 
words in $V_{\text{diff}}$ might have more occurance in the Whisper training data.
Conversely, Auto-AVSR and AV-HuBERT exhibit a $\Delta$ of approximately 10\%. This degradation indicates that these models might rely more heavily on the specific, expected lexicon.

A secondary factor contributing to this $\Delta$ is implicit overfitting. Because models are iteratively refined to achieve SoTA results on the standard LRS3 benchmark, researchers naturally select the checkpoints that perform best on this specific evaluation set. Consequently, the final models may become over-optimised for the benchmark's exact lexicon and environmental conditions, sacrificing broader generalisation for peak performance within a narrow testing environment.

\section{Dive into the Modalities}

To understand how these architectures process distinct streams of information, we tested their performance across different modality settings. 
We evaluated the models on the MV2LRS3 set using Audio-Visual (AV), Audio-Only (AO), and Video-Only (VO) inputs.

Table \ref{tab:modalities} reveals an unexpected trend: for several top-tier foundation models, adding visual data actually harms performance on the MV2LRS3 set. Llama-AVSR, Whisper-Flamingo, and USR all achieve lower WERs in the Audio-Only setting compared to the full Audio-Visual setting. This is not observed on LRS3. For example, Llama-AVSR scores 15.3\% on audio alone, but degrades to 16.5\% when the video stream is introduced. 
This gap appears to occur in both unified (AV, AO, VO in a single model) and separated (AV, AO, VO in separate models, task-specific) models, although the drop is slightly smaller in the unified models. 
Auto-AVSR appears to be an exception again here; its performance improves from 16.4\% (AO) to 14.0\% (AV), proving that it successfully gains advantage from using the visual modality.

Finally, the unimodal Video-Only (VO) results highlight the sheer difficulty of pure lip-reading on the MV2LRS3 set. All evaluated models struggle substantially without audio context. Notably, Llama-AVSR, despite being the best AV model on the original LRS3 benchmark, exhibits the highest error rate in the VO setting (81.5\%). 
This further confirms our earlier finding in the head yaw: the model's backend relies heavily on audio and linguistic context, and its visual encoder alone struggles to process unfamiliar visual environments.

\begin{table}[th]
\centering
\caption{WER (\%) of the audio-visual, audio-only, and video-only models on the LRS3 Test set and the MV2LRS3 set.}
\label{tab:modalities}
\setlength{\tabcolsep}{0.59\tabcolsep}
\begin{tabular}{lccccccc} 
\toprule
\multirow{2}{*}{\textbf{Model}} & \multirow{2}{*}{\begin{tabular}[c]{@{}c@{}}\textbf{Unified}\\\textbf{model}\end{tabular}} & \multicolumn{3}{c}{\textbf{LRS3}} & \multicolumn{3}{c}{\textbf{MV2LRS3}} \\ 
\cmidrule{3-5}\cmidrule(l){6-8}
 &  & \textbf{AV} & \textbf{AO} & \textbf{VO} & \textbf{AV} & \textbf{AO} & \textbf{VO} \\ 
\midrule
AV-HuBERT & $\checkmark$ & 1.5 & 2.0 & 34.1 & 23.5 & 23.6 & 65.1 \\
Auto-AVSR &  & 0.95 & 0.99 & 19.1 & 14.0 & 16.4 & 45.5 \\
USR & $\checkmark$ & 1.1 & 1.2 & 22.3 & 21.5 & 21.0 & 57.9 \\
Whisper-F. &  & 0.86 & 0.85 & – & 18.6 & 16.6 & – \\
Llama-AVSR &  & 0.77 & 0.81 & 24.0 & 16.5 & 15.3 & 81.5 \\
\bottomrule
\end{tabular}
\end{table}

\section{Dive into the Errors}
We here dive deep into the specific error: Substitutions (Sub), Deletions (Del), and Insertions (Ins) errors.  
As shown in Table \ref{tab:sdi}, the error profiles on the LRS3 Test set are extremely low and well-balanced across all three categories. However, the transition to the MV2LRS3 set reveals completely different error behaviours among the models.

The most prominent trend on the MV2LRS3 set is a severe spike in insertion errors for several models. 
AV-HuBERT, USR, and Whisper-Flamingo all exhibit insertion rates that are roughly double their substitution rates. 
This indicates that when these models encounter cross-corpus distributional shifts, they frequently hallucinate or generate extra words to compensate for missing context, rather than simply dropping unrecognised speech.

Llama-AVSR demonstrates a distinct contrast to this trend. While it maintains a relatively controlled insertion rate (6.4\%), it registers the highest deletion rate among all evaluated models (3.9\%). This suggests a different architectural behaviour: when faced with difficult or unfamiliar inputs on the MV2LRS3 set, Llama-AVSR is more conservative, dropping unrecognised words entirely rather than generating extra text.

Finally, Auto-AVSR maintains the most balanced error profile on the MV2LRS3 set. While its overall errors still increase compared to LRS3, it avoids the extreme insertion spikes seen in some other models, further explaining its top-ranking performance.

\begin{table}[th]
\centering
\caption{Substitutions (Sub), Deletions (Del), and Insertions (Ins) rates (\%) on the LRS3 Test and MV2LRS3 sets.}
\label{tab:sdi}
\begin{tabular}{lcccccc} 
\toprule
\multirow{2}{*}{\textbf{Model}} & \multicolumn{3}{c}{\textbf{LRS3}} & \multicolumn{3}{c}{\textbf{MV2LRS3}} \\ 
\cmidrule(lr){2-4} \cmidrule(lr){5-7}
& \textbf{Sub} & \textbf{Del} & \textbf{Ins} & \textbf{Sub} & \textbf{Del} & \textbf{Ins} \\ 
\midrule
AV-HuBERT & 0.9 & 0.2 & 0.4 & 7.1 & 1.6 & 13.1 \\
Auto-AVSR & 0.5 & 0.3 & 0.2 & 5.5 & 2.6 & 5.1 \\
USR & 0.7 & 0.2 & 0.2 & 6.2 & 1.8 & 12.0 \\
Whisper-F. & 0.4 & 0.3 & 0.2 & 4.1 & 1.8 & 11.2 \\
Llama-AVSR & 0.4 & 0.2 & 0.3 & 4.0 & 3.9 & 6.4 \\
\bottomrule
\end{tabular}
\end{table}

\section{Discussion}

\subsection{Does AVSR generalise beyond LRS3?}
The primary focus of this paper is the generalisation capability of current AVSR systems beyond the LRS3 benchmark.
Our empirical results reveal that performance on the controlled MV2LRS3 set universally collapses across all models, exposing a massive gap compared to the LRS3 Test set. Consequently, current AVSR models fail to generalise as robustly as image classifiers and VSR models \cite{recht2019imagenet, djilali2024vsr}. The considerably steeper slope of the AVSR linear fit highlights a magnified sensitivity to domain shifts, even when evaluated under these strictly matched conditions.

The steep slope of the linear fit in Equation (\ref{eq:linear}) serves as a critical warning regarding the current state of LRS3 benchmarking. The primary culprit for this phenomenon is performance saturation. Our experiments show that performance gains on the LRS3 benchmark are trending toward over-optimisation. Because state-of-the-art models are achieving WERs approaching the absolute noise floor of the dataset, the benchmark has become severely compressed. Consequently, the evaluation metric no longer scales linearly with genuine improvements in audio-visual representation.

\subsection{Hypothesising the Cause of the Performance Drop}

Our detailed attribute analysis, which isolated the impact of individual factors, reveals that duration is the primary driver of the observed performance degradation. We hypothesise that this steep drop is mainly driven by temporal overfitting. 
The profound sensitivity to duration suggests that current models are heavily over-optimised for the specific duration distribution and narrow context window of the LRS3 benchmark.

\subsection{Limitations and Unobserved Confounding Factors}
While our MV2LRS3 set methodology isolates the impact of seven primary acoustic and visual factors, we acknowledge the inherent limitations of this multidimensional matching. 
Despite our rigorous efforts to strictly align these factors with the LRS3 Test set distribution, we still observe a residual performance degradation. 
This remaining gap indicates the presence of unobserved confounding variables. 

We attribute this residual shift to two primary domains. 
First, other acoustic and visual factors that we did not consider, such as speaker accents \cite{feng2024towards}, speaking style (read speech or conversational) \cite{linke2025s} and visual noise like occlusion \cite{10888423}. 
We admit it is surprisingly hard to take all possible factors into account and construct a matched evaluation set. 
But this research is an initial step towards this goal. 

Second, and perhaps more critically, we must acknowledge the confounding role of non-transparent lexicon. 
Modern foundation models like Whisper and LLMs are trained on massive, proprietary, or undocumented corpora; hence, it is impossible to accurately assess their true baseline vocabularies. 
Consequently, we cannot isolate whether a specific transcription failure stems from challenging audio-visual geometries or simply from an unmeasurable out-of-vocabulary mismatch.

\subsection{Diminishing Multimodal Advantage}
Our cross-modality analysis reveals a severe architectural vulnerability: on the strictly matched MV2LRS3 set, almost all evaluated models achieve better performance on audio-only inference. 
This indicates that when confronted with the LRS3-like subset, the theoretical advantage of audio-visual fusion collapses. The visual modality does not merely fail to provide a supplementary benefit; it actively degrades overall system accuracy. 
This finding naturally raises a question: are AVSR models using the visual modality when testing outside of the LRS3 distribution, or are they treating the visual modality as noise?

The effective use of the visual modality is even more critical for AVSR in noisy, multi-speaker environments, such as the scenario used in the AVCocktail dataset \cite{nguyen25b_interspeech}. 
If current models cannot gain benefit from the visual modality under clean, matched conditions, it is questionable whether they are genuinely equipped to move on to a more challenging setting.

\section{Conclusion}

This paper systematically assesses the true generalisability of state-of-the-art AVSR systems beyond the standard LRS3 benchmark. 
To achieve this, we constructed the MV2LRS3 set: a controlled LRS3-like test set that strictly aligns with the demographic, acoustic and visual distributions of the LRS3 Test set. 
By evaluating five leading AVSR systems within this distribution-matched setting, we reveal a universal performance collapse. This provides empirical evidence that current models are severely over-optimised for the LRS3 benchmark, resulting in adaptive overfitting.

Furthermore, our attribute analysis isolates temporal duration as a primary driver of this failure. 
Our isolation of vocabulary impacts similarly exposes a profound implicit overfitting, where models demonstrate a severe lexical bias toward the expected LRS3 lexicon.
We also observe a collapse of the multimodal advantage: within a distribution-matched environment, the visual modality degrades overall system accuracy. 

Our experiments serve as a first step in assessing the generalisability of AVSR models, exposing the limitations and performance saturation inherent to the LRS3 benchmark.
This research establishes a structured framework for more comprehensive benchmarking practices, allowing more interpretable explanation beyond the overall WERs.
We publicly release the test sets and the associated metadata. We strongly urge the AVSR community to adopt this controlled benchmark to validate future architectures against potential adaptive overfitting. 
Ultimately, our detailed attribute analysis provides a blueprint for future data curation processes to be more considerate of the potential biasing factors. 

\section{Ethical Disclaimer}
Demographic metadata for this dataset was generated via automated extraction tools and represents algorithmic inference, not self-reported identity. 
We acknowledge the limitations of these off-the-shelf models, including potential algorithmic bias, uneven accuracy across demographics, and the reduction of complex traits into rigid categories. Researchers should utilise these labels with appropriate caution.

\section{Generative AI Use Disclosure}
Generative AI tools were used to improve the paper's grammar and clarity, as well as to assist in writing the code for the plots.

\section{Acknowledgements}
All data processing and experiments were conducted at Imperial College London and Trinity College Dublin. 
The work of Zhaofeng Lin and Naomi Harte was funded by the Research Ireland Centre for Research Training in Digitally-Enhanced Reality (d-real) under Grant No. 18/CRT/6224. 

\bibliographystyle{IEEEtran}
\bibliography{mybib}

\end{document}